\documentclass[aps,prl,twocolumn,superscriptaddress]{revtex4}
\usepackage{amsmath}
\usepackage{amsfonts}
\usepackage{amssymb}
\usepackage{verbatim}
\usepackage{amsthm}
\usepackage{graphicx}

\newcommand{\fig}{Fig.\ }

\begin{document}


\title{{Condensation in Temporally Correlated Zero-Range Dynamics}}
\author{Ori Hirschberg}
\author{David Mukamel}
\affiliation{Department of Physics of Complex Systems, Weizmann
Institute of Science, 76100 Rehovot, Israel}
\author{Gunter M. Sch{\"u}tz}
\affiliation{Institut f\"ur Festk\"orperforschung, Forschungszentrum
J\"ulich, 52425 J\"ulich, Germany}


\date{\today}


\begin{abstract}
Condensation phenomena in non-equilibrium systems have been modeled
by the zero-range process, which is a model of particles hopping
between boxes with Markovian dynamics. In many cases, memory effects
in the dynamics cannot be neglected. In an attempt to understand the
possible impact of temporal correlations on the condensate, we
introduce and study a process with non-Markovian zero-range
dynamics. We find that memory effects have significant impact on the
condensation scenario. Specifically, two main results are found: (1)
In mean-field dynamics, the steady state corresponds to that of a
Markovian ZRP, but with modified hopping rates which can affect
condensation, and (2) for nearest-neighbor hopping in one dimension,
the condensate occupies two adjacent sites on the lattice and drifts
with a finite velocity. The validity of these results in a more
general context is discussed.
\end{abstract}

 \maketitle

The Zero-Range Process (ZRP) is a paradigmatic model for mass
transfer in non-equilibrium systems
\cite{Spitzer1970,evanszrpreview}. In this model, particles hop
between boxes with hopping rates $u(n)$ which depend only on the
occupation number $n$ of the departing box. The steady state
distribution function of the box occupation numbers factorizes
into a product of single-box terms, which makes this model
amenable to theoretical studies. For rates which for large $n$ are
of the form
\begin{equation}\label{eq:hoppingrate}
u(n) = \gamma(1 + b/n),
\end{equation}
it  has been shown that the model exhibits a condensation
transition at high densities as long as $b>2$. At densities above
the condensation transition, one of the boxes is occupied by a
macroscopic number of particles, constituting a real-space
condensate. The parameter $\gamma$ sets the time scale of the
process.

The ZRP has been used to study condensation in shaken granular gases
where the shaken particles can move between compartments. Under
certain conditions condensation is observed, whereby most particles
accumulate in one compartment
\cite{vanderweele2001,torok2005granularzrp}. Similar condensation
phenomena may occur in network rewiring, resulting in a hub with a
macroscopic linking number
\cite{angeletal2005zrpnetworks,networksevolutionbook}. The ZRP has
also been used to model phase separation in driven diffusive systems
and traffic flow models, where, in the steady state, high and low
density regions coexist
\cite{kafrietal2002criterion,kafrietal2003phaseseperation,
levineetal2004trafficjams,kaupuzsetal2005zrptraffic}.
In modelling one dimensional driven diffusive systems, the length
of, say, a high density region in the driven system corresponds to
the occupation number of a box in the ZRP. The flow of particles
between high density domains corresponds to particle hopping in the
ZRP.

In reality, mass transfer processes usually constitute a
coarse-grained description of more complex microscopic dynamics.
This frequently results in temporally correlated dynamical
processes, which are not captured in the Markovian dynamics of the
ZRP \cite{chatterjeebarma2008}. It would thus be of interest to
explore the effect of temporal correlations on the collective
behavior of the ZRP, particularly on the occurrence of condensation.

In this Letter, we consider a non-Markovian ZRP, where the hopping
processes depend not only on the present occupation of a site but
also on jump events that have occurred in the past. We find that
this has significant impact on condensation. Specifically, two main
results are derived. The first is that memory effects on steady
state condensation can be captured by a Markovian ZRP with an
effective hopping parameter $b_\text{eff} \neq b$ which controls the
condensation transition. The second effect is that in the case of
asymmetric nearest-neighbor hopping on a ring of $L$ sites, the
condensate drifts with a finite velocity which scales as $1/L$.

The non-Markovian ZRP introduced in this work consists of $L$ boxes
$i$, each characterized by the particle number $n_i$ and an internal
``clock'' $\tau_i$. The total number of particles is $N$. The clock
proceeds irregularly in integer steps and is reset to zero each time
a particle jumps onto site $i$, thus keeping a memory of the
history of the process. A configuration of the system is then given
by the set $({\mathbf n}, \boldsymbol{\tau}) =
\{(n_i,\tau_i)\}_{i=1}^L$. The hopping rates out of site $i$ are
taken to depend both on the occupation of the site $n_i$ and on the
state of the clocks $\tau_i$, but not on the occupation of any other
site, in accordance with the general approach of the ZRP.

Here we consider the following dynamics: particles hop between
sites with rate $u(n,\tau)$. The internal clock on the target site
is reset to zero together with a jump. Independently of the jump
processes, at each site the internal clock is incremented by one
unit with a constant rate $c$. For a jump from $i$ to $j$, these
two processes can be schematically summarized by
\begin{align}\label{eq:modeldscrp}
(n_i,\tau_i),(n_{j},\tau_{j})&\xrightarrow{u(n_i,\tau_i)}(n_i\!-\!1,\tau_i),(n_{j}\!+\!1,\tau_j=0)
\nonumber \\
(n_i,\tau_i)&\xrightarrow[\phantom{u(n_i,\tau_i)}]{c}(n_i,\tau_i\!+\!1),
\end{align}
The particle jump process, when taken by itself, is non-Markovian,
since the rate of a jump depends on how much time has past since a
particle last hopped into the jump site. Choosing the target site
$j\neq i$ uniformly (i.e., considering a fully connected graph)
corresponds to mean-field (MF) dynamics, while restricting the
target site to $j=i+1$ corresponds to totally-asymmetric
nearest-neighbor hopping on a ring.

Notice that the full process of particle jumps and clock
increments together defines a Markovian process. Therefore, the
model may be implemented by a discrete-time Monte-Carlo version of
these dynamics with random sequential update which is defined as
follows: let $p_{\max} = \mbox{max}_{n,\tau} [u(n,\tau) +  c ]$.
For the Monte-Carlo update pick a pair of sites $(i,j)$ uniformly
and attempt to make one of the following changes: (i) move a
particle to the target site $j$ with probability
$u(n_i,\tau_i)/p_{\max}$ and reset the clock on the target site
$j$ to zero, or, (ii) increment the internal clock on the starting
site $i$ with probability $c/p_{\max}$. A total of $L$ consecutive
updates constitute one Monte-Carlo sweep.

Let us first consider the MF dynamics, whereby the target site $j$
is chosen uniformly. With this dynamics the steady state
distribution factorizes into a product of single site terms in the
thermodynamic limit. The single-site occupation and clock
probability in the steady state, $P(n,\tau)$, can then be found by
examining a single site with a ``mean-field'' incoming current $J$
generated by all other sites. This is equivalent to a solution of
a single site in the grand-canonical ensemble. The master equation
for the single site probability is
\begin{multline}\label{eq:mfmaster}
\frac{dP(n,\tau)}{dt} = -P(n,\tau)[J+c+u(n,\tau)] + {}\\
J P(n\!-\!1) \delta_{\tau,0} + c P(n,\tau\!-\!1) +
u(n\!+\!1,\tau)P(n\!+\!1,\tau),
\end{multline}
where the single site marginal distribution is defined by $P(n)
\equiv \sum_{\tau} P(n,\tau)$, and $J=\sum_{n,\tau}u(n,\tau)P(n,\tau)$ is the incoming current.
These equations are valid also for
$n=0$  and $\tau=0$ by defining $P(-1,\tau)=P(n,-1)=0$ and $u(0,\tau)=0$.

In the steady state, where $dP(n,\tau)/dt=0$, one obtains by summing
Eqs. (\ref{eq:mfmaster}) over $\tau$
\begin{equation}\label{eq:simplerec}
{\bar{u}}(n)P(n) = J P(n\!-\!1).
\end{equation}
Here $\bar{u}(n)$ is the
mean hopping rate out of a site with $n$ particles
\begin{equation}\label{eq:ubar}
\bar{u}(n) \equiv \frac{\sum_{\tau}
P(n,\tau)\,u(n,\tau)}{\sum_{\tau} P(n,\tau)}.
\end{equation}
Eq.\ (\ref{eq:simplerec}) expresses the balance between the
probabilities to hop into and out of a site with $n$ particles. The
marginal distribution $P(n)$ is therefore the same as that of a
Markovian ZRP \cite{andjel1982}, but with an \textit{effective}
hopping rate $\bar{u}(n)$. This gives the steady-state occupation
probability
\begin{equation}\label{eq:singlesiteprob}
P(n) = P(0)J^n\bar{f}(n)\quad \text{with} \quad \bar{f}(n) = \prod_{i=1}^{n}\bar{u}(i)^{-1}.
\end{equation}
Here $P(0)^{-1} \equiv 1+\sum_{n=1}^{\infty}J^n \bar{f}(n)$
ensures proper normalization of $P(n)$. The occurrence of
condensation is determined by the asymptotic behavior of $P(n)$.
For $P(n) \sim n^{-\alpha}$ condensation takes place for $\alpha >
2$ \cite{evanszrpreview}. In the Markovian case, ones has $\alpha
= b$ for hopping rates of the form (\ref{eq:hoppingrate}).

To analyze condensation in the non-Markovian case we proceed by considering,
for simplicity, hopping rates of the form
\begin{equation}\label{eq:onoffrates}
u(n,\tau) = \left\{
\begin{array}{ll} 0 & \tau =0 \quad\text{(``off'' state)}\\
u(n) & \tau \geq 1 \quad\text{(``on'' state).}
\end{array} \right.
\end{equation}
In this case, whenever a particle hops into a site the site is
switched to an ``off'' state in which no particles can hop out.
Only after its clock reaches $\tau=1$ the site is turned ``on''
again, in which case particles hop out with a rate
$u(n)$. This special case will be called the \textit{on-off
model}. In this case, the clock has in effect only two states: $\tau=0$ corresponding to ``off'', and $\tau \geq 1$
corresponding to ``on''. Correspondingly, the state of the site can be characterized by $P_\text{off}(n) = P(n,0)$ and
$P_\text{on}(n) = \sum_{\tau \geq 1}P(n,\tau)$.

The master equation for the stationary probability distribution
(\ref{eq:mfmaster}) is then
\begin{subequations}\label{eq:onoffmastereq}
\begin{align}
P_\text{off}(n)[J + c] &= J P(n\!-\!1) \label{eq:onoffmastereq1} \\
P_\text{on}(n)[J + u(n)] &= c P_\text{off}(n) +
P_\text{on}(n\!+\!1)u(n\!+\!1) \label{eq:onoffmastereq2} .
\end{align}
\end{subequations} First we study the probability $P_\text{off}$ to
find a site in the ``off'' state. By summing over $n$, Eq.\
(\ref{eq:onoffmastereq1}) yields
\begin{equation}\label{eq:poff}
P_{\text{off}} = \frac{J}{c\!+\!J}.
\end{equation}
Further, from Eqs.\ (\ref{eq:simplerec}),
(\ref{eq:onoffmastereq1}) and \eqref{eq:poff} we find
\begin{equation}\label{eq:onofffullprob1}
P_{\text {off}}(n) = \frac{\bar{u}(n)P(n)}{J+c} = \frac{P_{\text
{off}}}{J}\bar{u}(n)P(n).
\end{equation}
In addition, for $u(n,\tau)$ of the form \eqref{eq:onoffrates},
Eq.\ \eqref{eq:ubar} gives
\begin{equation}\label{eq:onofffullprob2}
P_{\text {on}}(n) = \frac{\bar{u}(n)}{u(n)}P(n).
\end{equation}

As discussed above, the occurrence of condensation depends only on
the asymptotic behavior of the effective hopping rate,
$\bar{u}(n)$. The effective hopping rate follows from
\eqref{eq:onofffullprob1} and \eqref{eq:onofffullprob2} using
$P_\text{off}(n) + P_\text{on}(n) = P(n)$, and is given by
\begin{equation}\label{eq:onoffubar}
\frac{1}{\bar{u}(n)} = \frac{P_{\text{off}}}{J}+\frac{1}{u(n)}.
\end{equation}
This states that the mean time between hops from a site with $n$ particles
is equal to the mean time this site is in an ``off'' state plus
the time it takes a
particle to hop out once the system is already ``on''.

In the case of $u(n)$ of the form (\ref{eq:hoppingrate}), it can
be seen from (\ref{eq:onoffubar}) that to leading order in $1/n$
the effective hopping rates are are again of the form
(\ref{eq:hoppingrate}), given by
\begin{equation}\label{eq:onoffubarfinal}
\bar{u}(n) \sim
\frac{c+J}{c+J+1}\bigg(1+\frac{b_{\text{eff}}}{n}\bigg),
\end{equation}
with
\begin{equation}\label{eq:beff}
b_{\text{eff}}=\frac{c+J}{c+J+1}\, b \, < \, b.
\end{equation}
From (\ref{eq:singlesiteprob}), we then have
\begin{equation}
P(n) \sim  \biggl[\frac{J(J+c+1)}{J+c}\biggr]^n n^{-b_\text{eff}}.
\end{equation}

To analyze the condensation transition we note that for
${J(J+c+1)}/{(J+c)}<1$, the distribution decays exponentially with
$n$ and the system is in a subcritical homogeneous phase. For a
critical current given by $J_c = {(c+J_c)}/{(c+J_c+1)}$, the decay
is algebraic, and one has condensation for $b_\text{eff}>2$. For
the critical current we find
\begin{equation}\label{eq:onoffJc1b}
J_c = \frac{c}{2}\bigg(\sqrt{1+\frac{4}{c}}-1\bigg),
\end{equation}
which allows us to write
\begin{equation}
b_\text{eff} = J_c\cdot b.
\end{equation}
Therefore, condensation takes place for hopping parameter $b >
\frac{4}{c}\left(\sqrt{1+4/c}-1\right)^{-1}$ which is larger than 2,
in contrast with
the Markovian case for which the critical value for condensation
is $b=2$. This means that for densities $\rho=N/L$
above a critical density $\rho_c$, the average
number of particles in the condensate
is $N_\text{cond}=L(\rho-\rho_c)$. All other sites
have an average ``background'' density
$\rho_c$.

\begin{figure}
  \includegraphics[width=0.4\textwidth]{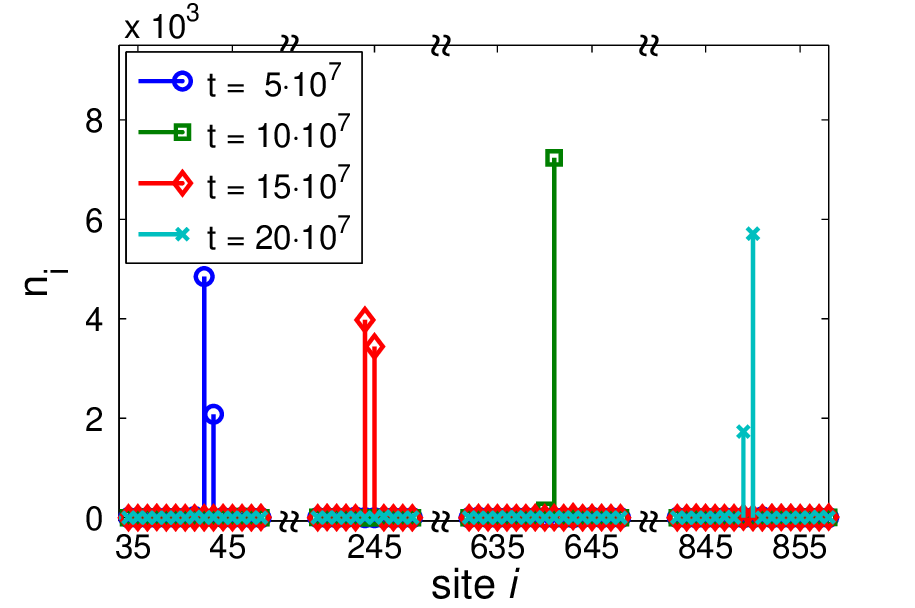}
  \caption{ \label{fig:ringcondensatesnapshot}{Snapshots of the on-off model with
 totally-asymmetric
  nearest-neighbor hopping on a ring, showing the occupation numbers $n_i$ at and in the
vicinity of the condensate at four points in time.
  Here, $L=1000$ sites, $\rho = 10$
  $b = 5.5$ and $c=1$.
  The condensate occupies two sites
  and drifts with a constant mean velocity.
  }}
\end{figure}

We now turn to a ring geometry with nearest neighbor hopping and
fully asymmetric dynamics. In this case, the stationary
distribution does not factorize. In order to get insight into the
condensation mechanism in this case we carried out numerical
simulations of the on-off model with $u(n)$ given by Eq.\
(\ref{eq:hoppingrate}).

Simulations in the condensed phase indicate that the condensate
drifts with a finite velocity. In addition, we find that in
contrast to previously known condensation phenomena, the
condensate typically occupies two adjacent sites $i$ and $i+1$.
These observations can be explained by a more detailed microscopic
investigation of the dynamics. It turns out that the drift takes
place via a ``slinky'' motion where particles hop from site $i$ to
$i+1$ at a rate $u(n_i)$ which is approximately constant at large
$n_i$, leaving site $i+1$ in a predominantly off-state. Thus
particles accumulate at site $i+1$ until site $i$ is emptied,
giving the clock at $i+1$ the chance to reach the on-state for
durations of time sufficiently long to allow particles to escape.
Then particles start to hop from site $i+1$ to site $i+2$ in the
same fashion.

This mechanism suggests that the drift velocity $v_\text{cond}$ is
inversely proportional to the number of particles in the
condensate $N_\text{cond}$, i.e.,
\begin{equation}
v_\text{cond}^{-1}\sim N-N_c = L(\rho-\rho_c).
\end{equation}
In the thermodynamic limit, the velocity of the condensate vanishes.
Superimposed on this motion, the condensate can melt and reappear at
some other site of the lattice, similar to what happens in the
Markovian case. This happens on a characteristic time which scales
with the system size to a power larger than 2
\cite{grosskinskyetal2003zrpcondensation,godrecheluck2005condensate,beltramlandim2008}.

\begin{figure}
\includegraphics[width=0.4\textwidth]{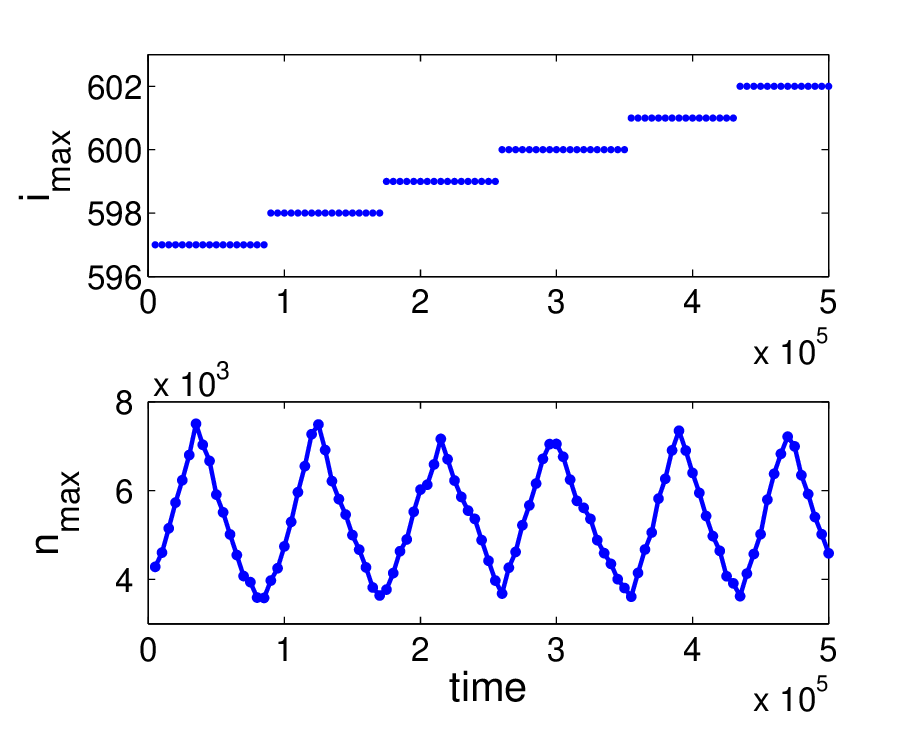}
  \caption{{The position of the most occupied site
  $i_{\max}$ and its occupation number $n_{\max}$ as a
  function of time, on a ring of size $L=1000$ with totally-asymmetric
  nearest-neighbor hopping. Simulation parameters: $\rho = 10$
  $b = 5.5$ and $c=1$.
  }}
  \label{fig:ringcondensatedrift}
\end{figure}

In Fig.\ \ref{fig:ringcondensatesnapshot} we present snapshots at
different times of the on-off model with totally asymmetric
nearest neighbor hopping for $L=1000$. One clearly sees that the condensate
occupies two adjacent sites with varying relative occupation of the two
sites, consistent with the slinky motion described above. The drift
of the condensate is evident in the figure. In order to
demonstrate the slinky motion in more detail, we present in Fig.\
\ref{fig:ringcondensatedrift} a plot showing the position of the most
occupied site $i_\text{max}$ and its occupation number,
$n_\text{max}$, as a function of time. The occupation
number $n_\text{max}$ oscillates in
time with approximately constant frequency. Typically it decreases linearly
until it reaches its minimal value, when $i_\text{max}$ increases
by 1 and $n_\text{max}$ starts increasing.

\begin{figure}
  \includegraphics[width=0.4\textwidth]{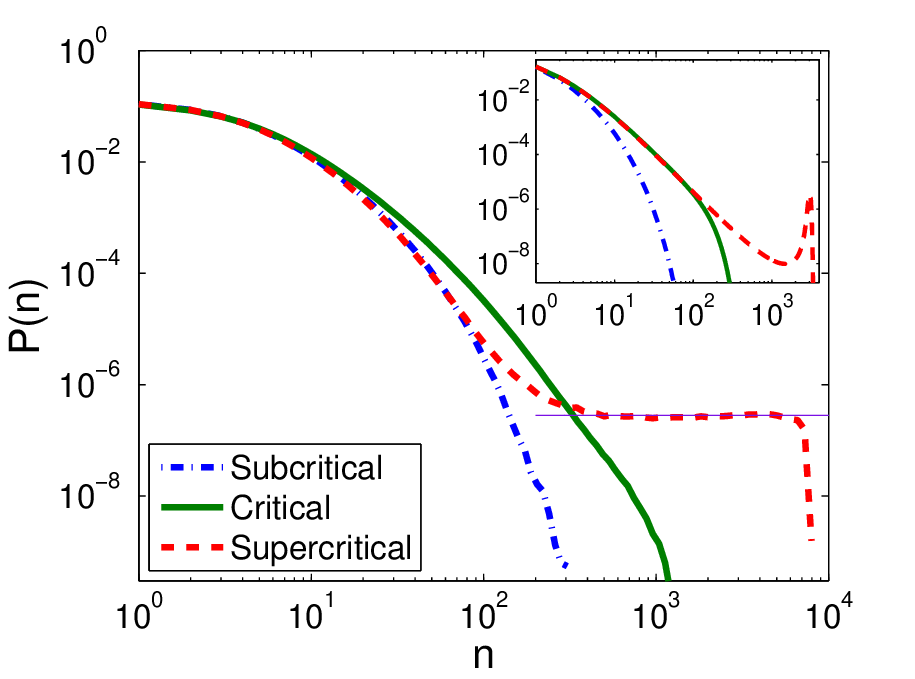}
  \caption{{The occupation probability $P(n)$ of a single-site in a ring with
  totally-asymmetric nearest-neighbor hopping and different densities,
  as obtained from Monte-Carlo simulations. Here $L=1000$, and curves
  for subcritical density ($\rho = 3$), supercritical density
  ($\rho = 10$) and at
  the critical region ($\rho=4.1$) are presented. The horizontal
  line indicates $P_\text{plateau}$ of \eqref{eq:ringpcond}, where
  $\rho_c$ was obtained from the simulation.
  The inset shows a similar plot of $P(n)$ for a Markovian ZRP
  of $L=1000$ sites with $b=5$, in the subcritical ($\rho = 0.5$) critical
  ($\rho = 1$) and supercritical ($\rho = 4$) phases.
  }}
  \label{fig:ringpofn}
\end{figure}

In Fig.\ \ref{fig:ringpofn} we present the single-site occupation
probability distribution $P(n)$ for various densities. At high
densities the distribution exhibits a plateau which reflects the
particle distribution among the two sites which constitute the
condensate. This is in contrast with a Markovian ZRP where the
condensate is supported by a single site, which results in a sharp
peak in $P(n)$. The value of $P(n)$ at the plateau in the
non-Markovian case may be estimated for $\rho$ above the critical
density and large $L$ using the slinky motion of the condensate.
The probability that a given site carries the condensate is $2/L$,
and in such a site there is an approximately uniform probability
to find any occupation $0<n<N_\text{cond}=L(\rho-\rho_c)$. Thus,
\begin{equation}\label{eq:ringpcond}
P_\text{plateau} \sim \frac{2}{L}\cdot\frac{1}{L(\rho-\rho_c)}.
\end{equation}
This estimate agrees well with the plateau value in \fig
\ref{fig:ringpofn}. For small densities,  $P(n)$ decays
exponentially, indicating the absence of a condensate. For the
systems size studied in this figure, the distribution at small
values of $n$ does not allow to extract a power law decay as
expected for the condensation transition. At density $\rho = 4.1$
there is a range of $n$ for which $P(n)$ seems to follow a power
law with $b_\text{eff} \approx 4$. This value differs
significantly from $b = 5.5$, which is the expected power for the
Markovian ZRP.

So far, we discussed in detail a simple on-off model which has
been used to demonstrate the effect of non-Markovian dynamics on
the steady state distribution and on condensation. For MF
dynamics, the asymptotic behavior of the effective rates can be
obtained for the more general class of models with rates of the
form $u(n,\tau) = u(n)v(\tau)$. Generally, $b_\text{eff}$ may be
larger or smaller than the ``bare'' value of $b$. Interestingly,
one can introduce an on-off model with nearest-neighbor hopping
for which the stationary distribution factorizes strictly even on
finite lattices and where the effective hopping rates can be
computed exactly. In this model particles jump with rates
(\ref{eq:onoffrates}) but the advancement of the clock depends on
the clock states of neighboring sites \cite{guntersmodel}.

The findings of this work suggest that the temporal correlations may
significantly alter the condensation transition and the nature of the
condensate in general driven systems. It would be of interest to study
broader classes of temporal correlations in order to explore other
scenarios of non-equilibrium condensation.

We thank M.\ R.\ Evans for useful comments. The support of the
Israel Science Foundation (ISF) and the Albert Einstein Minerva
Center for Theoretical Physics is gratefully acknowledged. This work
was carried out in part while G.\ M.\ S.\ was Weston Visiting
Professor at the Weizmann Institute of Science.


\bibliography{paperbib}

\end{document}